\documentclass[journal=ancac3,manuscript=article, layout=column]{achemso}

\usepackage[version=3]{mhchem} 

\usepackage{graphicx}
\usepackage{bm}
\usepackage{color}
\usepackage{upgreek}
\usepackage{amssymb}
\usepackage{amsmath}
\usepackage{booktabs}
\usepackage{multirow}
\usepackage{siunitx}

\newcommand{\sst}[1]{\scriptscriptstyle{#1}}

\newcommand{\DelDO}{\Delta_{\mbox{\tiny{DO}}}}
\newcommand{\DelDP}{\Delta_{\mbox{\tiny{DP}}}}
\newcommand{\vecu}{\mbox{\boldmath$u$}}
\newcommand{\vecin}{\mbox{\boldmath$u$}_{\mbox{\tiny{inlet}}}}
\newcommand{\vecp}{\mbox{\boldmath$u$}_{\mbox{\tiny{p}}}}
\newcommand{\vecdp}{{\mbox{\boldmath$u$}}_{\mbox{\tiny{DP}}}}
\newcommand{\vecdo}{{\mbox{\boldmath$u$}}_{\mbox{\tiny{DO}}}}
\newcommand{\xDOPS}{x_{\scriptscriptstyle \textrm{DOPS} }}
\newcommand{\cH}{c_{\scriptscriptstyle H}}
\newcommand{\cL}{c_{\scriptscriptstyle L}}
\newcommand{\bIH}{\bar{I}_{\scriptscriptstyle H}}
\newcommand{\bIL}{\bar{I}_{\scriptscriptstyle L}}
\newcommand{\dDLS}{d_{\scriptscriptstyle DLS}}
\newcommand{\dMF}{d_{\scriptscriptstyle MF}}

\newcommand{\Gammadp}{\Gamma_{\mbox{\tiny DP}}}
\newcommand{\Gammado}{\Gamma_{\mbox{\tiny DO}}}
\newcommand{\Nab}{\mbox{\boldmath$\nabla$}}

\newcommand{\PreserveBackslash}[1]{\let\temp=\\#1\let\\=\temp}
\newcolumntype{C}[1]{>{\PreserveBackslash\centering}p{#1}}
\newcolumntype{R}[1]{>{\PreserveBackslash\raggedleft}p{#1}}
\newcolumntype{L}[1]{>{\PreserveBackslash\raggedright}p{#1}}

\author{Adnan Chakra}
\affiliation{Department of Chemical Engineering, Loughborough University, Loughborough, LE11 3TU, United Kingdom}
\alsoaffiliation{Department of Chemistry, University College London, London, WCH1 0AJ, United Kingdom}
\author{Naval Singh}
\affiliation{Manchester Centre for Nonlinear Dynamics, Department of Physics and Astronomy, University of Manchester, Manchester M13 9PL, United Kingdom}
\author{Goran T. Vladisavljevi\'{c}}
\affiliation{Department of Chemical Engineering, Loughborough University, Loughborough, LE11 3TU, United Kingdom}
\author{Fran\c cois Nadal}
\affiliation{Commissariat \`{a} l'\'{E}nergie Atomique, BP2, 33114, Le Barp, France}
\author{C\'{e}cile Cottin-Bizonne}
\author{Christophe Pirat}
\affiliation{Institut Lumi\`{e}re Mati\`{e}re, UMR5306 Universit\'{e} Claude Bernard Lyon 1
- CNRS, Universit\'{e} de Lyon, Villeurbanne Cedex, 69622, France}
\author{Guido Bolognesi}
\affiliation{Department of Chemistry, University College London, London, WCH1 0AJ, United Kingdom}
\alsoaffiliation{Department of Chemical Engineering, Loughborough University, Loughborough, LE11 3TU, United Kingdom}
\email{g.bolognesi@ucl.ac.uk}

\title{
Continuous manipulation and characterization of colloidal beads and 
liposomes via diffusiophoresis
in single- and double-junction microchannels
}

\begin{document}


\begin{tocentry}


%
%
%

\includegraphics[scale=1]{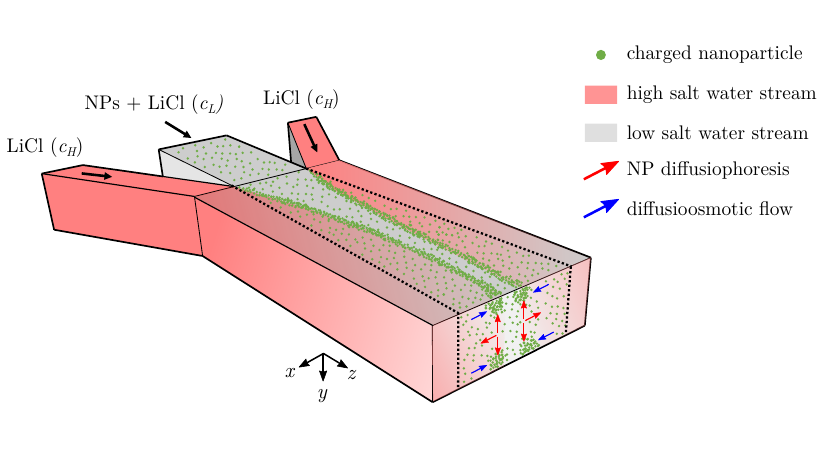} 

\end{tocentry}

\begin{abstract}

We reveal an unreported physical mechanism that enables the pre-concentration,
sorting and characterization of charged polystyrene nanobeads  and liposomes dispersed in a
continuous flow within a straight micron-sized channel. Initially, a single
$\Psi$-junction microfluidic chip is used to generate a steady-state salt
concentration gradient in the direction perpendicular to the flow. As a result,
fluorescent nanobeads dispersed in the electrolyte solutions accumulate into symmetric
regions of the channel, appearing as two distinct symmetric stripes when
the channel is observed from top via epi-fluorescence microscopy. 
Depending on the electrolyte flow configuration and, thus,
the direction of the salt gradient field,
the fluorescent stripes get closer to or apart from each other
as the distance from the inlet increases.
Our numerical and experimental analysis shows that,
although diffusiophoresis and hydrodynamic effects are involved in the
accumulation process, diffusioosmosis plays a crucial role in the observed
particles dynamics. In addition, we developed a proof-of-concept double 
$\Psi$-junction microfluidic device which exploits this accumulation mechanism for the size-based
separation and size detection of nanobeads as well as for the
measurement of zeta potential and charged lipid composition of liposomes
 under continuous flow settings. 
This device is also used to investigate the effect
of fluid-like or gel-like states of the lipid membranes on the liposome diffusiophoretic response.
The proposed strategy for solute-driven manipulation and characterization of colloids have great potential for
microfluidic bio-analytical testing applications, including bioparticle
pre-concentration, sorting, sensing and analysis.

\end{abstract}



\section{Introduction}
Synthetic and natural nanoparticles are ubiquitous in a wide range of chemical~\cite{moshfegh2009nanoparticle},
bioanalytical~\cite{couvreur2013nanoparticles,penn2003nanoparticles},
biomedical~\cite{mitchell2021engineering,stremersch2016therapeutic,chen2016nanochemistry}
and environmental~\cite{liu2006nanoparticles} applications.
Control over nanoparticle motion is
an important aspect for many of these applications, especially for bio-analysis, drug delivery, diagnostics and
environmental monitoring
for which particle filtration, pre-concentration, directed delivery and purification 
are often required~\cite{li2017progress,xie2020microfluidic,pamme2007continuous}. 
Nanoparticle characterization in terms of size and surface properties -- such
as chemical composition and charge -- is also key to several nanoparticle technologies.
For example, particle size and zeta potential play a crucial role in 
many biological processes, such as membrane-binding affinity,
cellular uptake and cytotoxicity~\cite{honary2013effect, miller1998,contini2018nanoparticle,contini2020size}. Size, surface charge and
surface composition  also dictate  the nanoparticles' ability to penetrate through natural barriers, such as 
extracellular matrices~\cite{lenzini2020matrix,tomasetti2016influence} and the blood brain 
barrier~\cite{huang2022potential}. 
Consequently, control over particle size and surface properties is essential for precision drug delivery applications,
in which these properties are tuned 
to increase circulation time~\cite{torchilin2005, yuan2012}, achieve high
therapeutic efficacy~\cite{ibrahim2005,gradishar2006} and reduce
toxicity~\cite{blanco2015}.
As a further example, extracellular vesicles -- i.e., lipid vescicle naturally released and internalized by cells --
has attracted much attention because of their potential as powerful therapeutic and diagnostic tools~\cite{fais2016evidence}.
Their size, surface charge and biochemical composition reflect their biogenesis
and determine the cellular uptake pathways used for intercellular communication~\cite{midekessa2020zeta}.
Thus, the characterization of the size and surface properties of these lipid vesicles is necessary to elucidate the 
many physiological and pathological processes they are involved with~\cite{yanez2015biological}, and
to exploit their potential as drug carrier and disease biomarkers.

In the last two decades, there has been a growing interest in microfluidic
strategies for both particle manipulation~\cite{zhang2020concise,sajeesh2014particle} and
characterization~\cite{gholizadeh2017microfluidic,yang2018review}.
This is due to the
many advantages offered by microfluidic technologies compared to
their conventional counterparts, including cost-effectiveness, reduced consumption
of samples and reagents, high precision, portability and capability to perform
multiplexed analysis and continuous flow processing~\cite{samiei2016, sackmann2014, rodriguez2021}. 
In this context, diffusiophoresis -- namely, the phoretic transport of colloidal particles
induced by a solute concentration gradient -- has emerged as a valuable
tool for particle manipulation and characterization in microfluidic chips~\cite{shin2020diffusiophoretic,shim2022diffusiophoresis},
and many devices have been proposed to exploit it for
targeted delivery~\cite{shin2016size,tan2021two}, focusing~\cite{Abecassis2008d}, trapping~\cite{singh2020reversible,singh2022enhanced}, 
accumulation~\cite{shin2017accumulation}, sorting, filtration 
and separation~\cite{shimokusu2019colloid,shin2017co,ault2018diffusiophoresis,ha2019dynamic,seo2020continuous}.
Furthermore, since the diffusiophoresis mobility of colloids can depend on particle 
size~\cite{shin2016size,rasmussen2020size, shimokusu2019colloid} and surface charge
\cite{prieve1984motion, velegol2016origins, shin2017low}, this transport mechanism
has been exploited also for the detection and characterization these particle properties.
A low-cost zeta potentiometry microchip was developed by relying on 
the fluid and particle motion induced within dead-end microchannels
via transient salt concentration gradients~\cite{shin2017low}.
This microfluidic device has a simple geometry, and it is cheap and easy to fabricate. 
However, it allows only for batch (i.e. non-continuous) measurements,
due to the transient nature of the salt concentration gradient and the
need for regular flushing of the dead-end pores to replace the sample and prevent clogging.

Imposing a salt concentration gradient in a microchip also results in a diffusioosmotic
slip velocity at the charged walls of the fluidic
channels~\cite{anderson1989colloid,marbach2019osmosis}.  Such a slip velocity
is typically weak and has usually negligible effects in pressure-driven flows
within open micron-sized channels. Conversely, diffusioosmotic effects can
become significant in dead-end channels~\cite{kar2015enhanced,shin2016size} or
in highly confined flows within nanotubes~\cite{siria2013giant} and
nanochannels~\cite{lee2014osmotic}.  Recently, diffusiophoresis and
diffusioosmosis have been jointly exploited in a cleverly designed microfluidic platform
for the separation and characterization of liposomes and extracellular
vesicles~\cite{rasmussen2020size}.  In this device, a shallow tapered
open-ended nanochannel is exposed to a steady salt concentration gradient and
this leads to the size- and charge-dependent accumulation of lipid vesicles
nearby the region where the diffusioosmotic flow velocity and the particle
diffusiophoretic velocity balance each other.  This microfluidic system enables
the pre-concentration of lipid vesicles as well as the accurate measurements of their
diameter and zeta potential. Although this is the first significant exploitation of diffusioosmotic flows
for particle characterization, the sample analysis is again performed in batches, with a small fraction
of the sample injected into the device being analyzed at any given time. 
Moreover, the use of nanochannels makes the device fabrication process
more complex and expensive, because it requires costly cleanroom fabrication procedures such as
reactive ion etching techniques~\cite{lee2014osmotic,marbach2019osmosis,rasmussen2020size}. 
Nanochannel devices are also more vulnerable to obstruction and clogging,
compared to microchannel chips.

To address these limitations, we propose a novel microfluidic strategy for the manipulation
and characterization of colloidal particles via solute concentration gradients
in continuous flow, low-cost and easy to fabricate microfluidic devices.
First, we report and investigate a newly observed particle focusing mechanism occurring in a single $\Psi$-junction microchannel device.
The intertwined effects of particle diffusiophoresis and flow
diffusioosmosis, induced by steady salt concentration gradients, 
enable the accumulation of colloidal beads and liposomes at multiple focusing regions
nearby the charged walls of the microchannels.
Notably, unlike previous studies, here 
the diffusiophoresis and diffusioosmosis effects do not compete one against the other by pushing
particles along parallel and opposite directions, but, instead, they
act synergistically by moving particles along two perpendicular directions.
This allows for the first time the continuous and high-throughput manipulation of colloids in 
an open-ended micron-sized channel via diffusioosmotic flows
without recurring to dead-end or nano-sized channel geometries, therefore reducing the risk
of device clogging and avoiding the need of expensive device fabrication procedures.
To exploit this novel mechanism, we introduce a new double-junction device that can be
used for the continuous separation and characterization of
nanoparticles based on their size or zeta potential.
Furthermore, we investigate how the chemical composition 
and lipid phase of the membrane affect the liposome diffusiophoretic response
and establish a relationship between the latter and charged lipid content of the membrane.


\section{RESULTS AND DISCUSSIONS}

\subsection{Particle focusing in single junction devices}

\begin{figure}[t!] \begin{center}
\includegraphics[width=0.90\columnwidth]{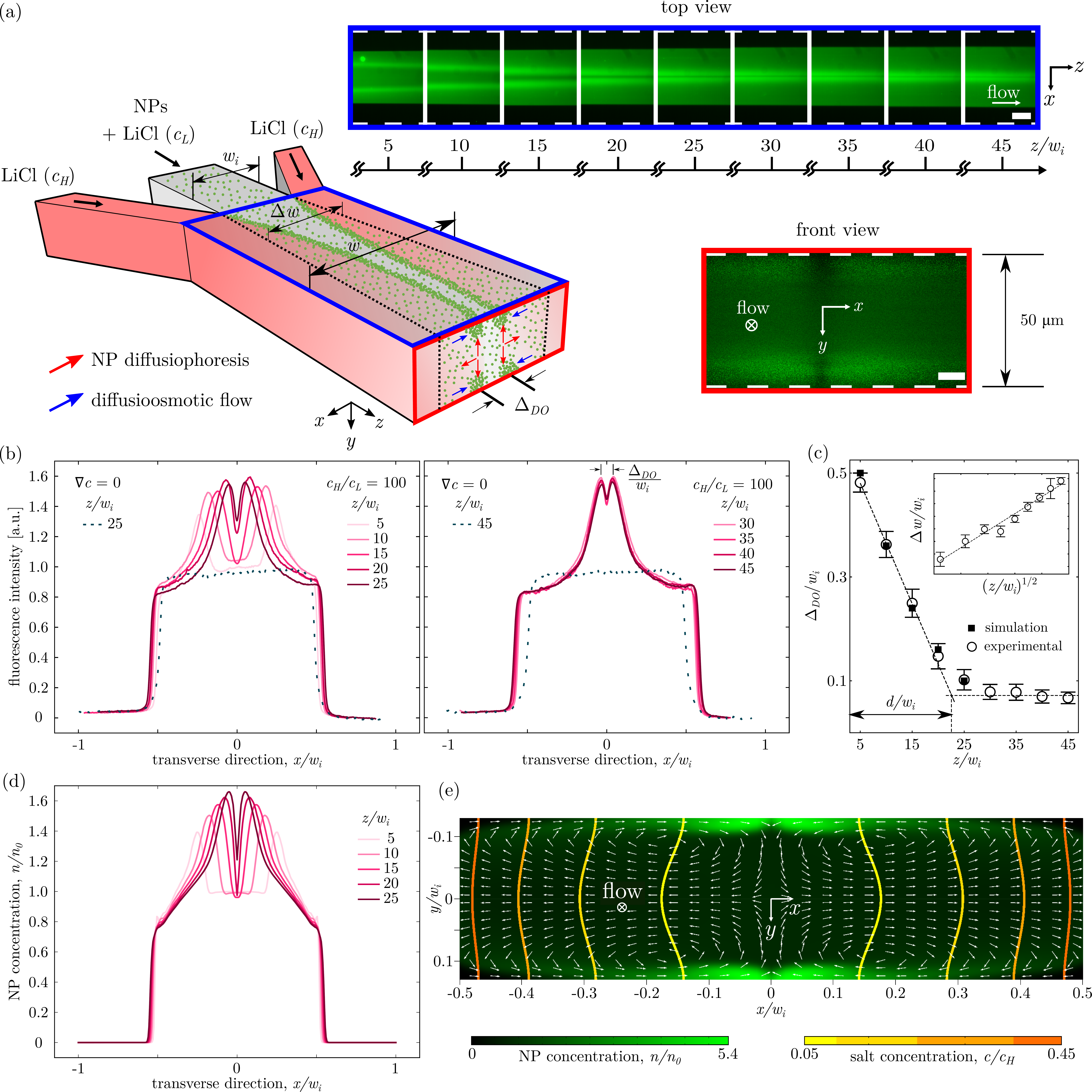} 
\end{center} 
\caption{
\footnotesize
(a)
Schematic and micrographs of the single $\Psi$-junction device.
Outer streams: LiCl solution at high concentration ($\cH$). 
Inner stream: LiCl solution at low
concentration ($\cL$) seeded with fluorescent ($d$ = $216.5\pm 0.9~$nm) carboxylate polystyrene nanoparticles (NPs), shown as green dots.
The decay of the distance $\DelDO$ between the accumulation peaks with the
distance from the junction ($z$) is visible in the top view epi-fluorescence micrographs (blue
rectangle). The micrographs were acquired with a focal plane located nearby
the bottom glass wall of the device. 
The four accumulation regions in the transverse  $(x,y)$-plane are visible in
the confocal image of the channel cross-section at a distance $z/w_i=25$ from the junction (front view, red
rectangle). (b) Experimental intensity profiles along the transverse
$x$-direction with a salt gradient (solid lines) and
without a salt gradient (dashed lines), at increasing distances $z/w_i$.
(c) Experimental and numerical peak separation distance $\DelDO$ as a function of the
distance $z/w_i$. 
Inset shows the profile of the colloidal band width, $\Delta w$, as a function of the distance $z/w_i$. 
(d) $y$-averaged NP concentration profiles along the transverse $x$-direction at increasing
distances $z/w_i$, predicted by numerical simulations. 
(e) Numerical map of the channel cross-section at a distance $z/w_i=25$
showing the NP concentration field, the salt concentration isolines and the total particle
velocity field, $\vecp=\vecu+\vecdp$, streamlines (white arrows)
with diffusioosmotic slip velocity, $\vecdo$, at the channel walls.
\label{fig:Fig1}} 
\end{figure}

A $\Psi$-junction microfluidic device was used to generate a steady-state
gradient of salt concentration (referred to as $c$ in the following) by pumping
a low concentration ($\cL = 0.1~$mM) LiCl aqueous solution in the inner channel
and a high concentration ($\cH = 10~$mM) LiCl aqueous solution in the outer
channels (Fig.\,\ref{fig:Fig1}a). Outer and inner streams had a constant flow
rate of $3.65~\upmu$L/min. 
A similar flow configuration was adopted in previous
studies~\cite{singh2022enhanced,
singh2020reversible,singh2022composite,Abecassis2008d,Abecassis2009b} to
investigate the dynamics of charged colloids under steady-state salt gradients
and continuous flow conditions.

In a first set of experiments, carboxylate polystyrene nanoparticles (NPs),
$216.5\pm 0.9~$nm in diameter (Invitrogen, USA) and $\zeta=-54.9\pm0.7$~mV, were dispersed in the inner flow only,
thus leading to the formation of a fluorescent stream of colloidal solution –
hereby referred to as colloidal band – within the central region of the
channel. The boundaries of the colloidal band are represented by the black
dotted lines in the 3D schematic of the device in Fig.\,\ref{fig:Fig1}a. 
A $x,y,z$-axis reference system is introduced as shown in the figure. The origin of $z$-axis is
located at the junction, whereas the origins of the $x$ and $y$ axis are located at the mid-points 
along the channel width and depth, respectively.
The parallel streams of electrolyte solutions at different salt concentrations
generated a salinity gradient in the direction transverse to the flow (red
arrows parallel to $x$-direction in Fig.\,\ref{fig:Fig1}a), which caused the broadening of the colloidal
band due to the diffusiophoresis migration of  charged nanoparticles towards
higher salt concentration regions.  In agreement with previous
studies~\cite{Abecassis2008d,Abecassis2009b}, the particle transverse profiles
showed an enhanced diffusive dynamics in the transverse direction and the
colloidal band width, $\Delta w$, increased linearly with the square root of
the longitudinal distance $z$ from the junction (inset in
Fig.\,\ref{fig:Fig1}c). As discussed by Abecassis and co-workers, the enhanced
colloid diffusivity in the transverse direction is caused by the coupling of
the diffusiophoretic migration of particles with the underlying salt diffusion
process~\cite{Abecassis2008d}.

The chemical gradient field also prompted the appearance of two
distinct peaks of focused nanoparticles (when looking at the channel from above), as showed
by the epi-fluorescence images in the blue rectangle of Fig.\,\ref{fig:Fig1}a.
The peaks were separated from each other by a distance $\DelDO$, initially equal to the inner channel
width $w_i$. The value of $\DelDO$ rapidly decreased with $z$ over a typical distance $d\simeq 23\,w_i$
and eventually plateaued at a constant \emph{saturation} value (Fig.\,\ref{fig:Fig1}c). 
The decay distance $d$ was estimated by 
intersecting two straight lines fitting the $\DelDO$ versus $z$ plot for $z/w_i \rightarrow 0$ and $ z/w_i \rightarrow 45$, respectively.
Particle peaks did not form when no salt concentration gradient was imposed (Fig.\,S1b,c).
The particle focusing effect is quantified by plotting the fluorescence intensity profile
against the transverse $x$-direction (perpendicular to the direction of the flow) at different
distances, $z/w_i$, downstream the junction (Fig.\,\ref{fig:Fig1}b).
The profiles are 
normalized with respect to the intensity of
the colloidal solution with particle concentration, $n_0$, injected in the inner channel of the device. 
Note that the epi-fluorescence micrographs are generated from the convolution of the particle fluorescence intensity
with the microscope point-spread function~\cite{gu2000advanced}. Thus, the intensity profiles of Fig.\,\ref{fig:Fig1}b
are the result of an integration of the particle fluorescence intensity
over an optical window, whose characteristic size in the vertical $y$-direction 
is given by the depth of field of the optical system (ca. $10\,\upmu$m).
For the micrographs in Fig.\,\ref{fig:Fig1}a, the mid-point of this optical window, 
which coincides with the focal plane of
the microscope objective, 
was located nearby the bottom glass wall of the device.
Note that epi-fluorescence images were also acquired with the focal plane located nearby the top PDMS wall and at 
the mid-point along the depth of the microchannel, but no significant change in the micrographs and
corresponding intensity profiles could be detect.
From the intensity profiles in Fig.\,\ref{fig:Fig1}b,  we can observe a clear
effect of focusing of the nanoparticles (solid lines) when compared to the case
without salinity gradient (dashed lines). A confocal scan of the $(x,y)$-plane
confirms that under salinity gradient conditions, charged particles were
accumulating at the top and bottom walls of the device in four symmetrical
positions (red rectangle in Fig.\ref{fig:Fig1}a).  In the absence of a salinity
gradient, the particle distribution profile in the same plane was uniform
(Fig.\,S1c), therefore the observed particle dynamics was  driven by the salt
contrast generated in the microchannel.

\begin{figure}[t!]
\begin{center}
\includegraphics[width=\columnwidth]{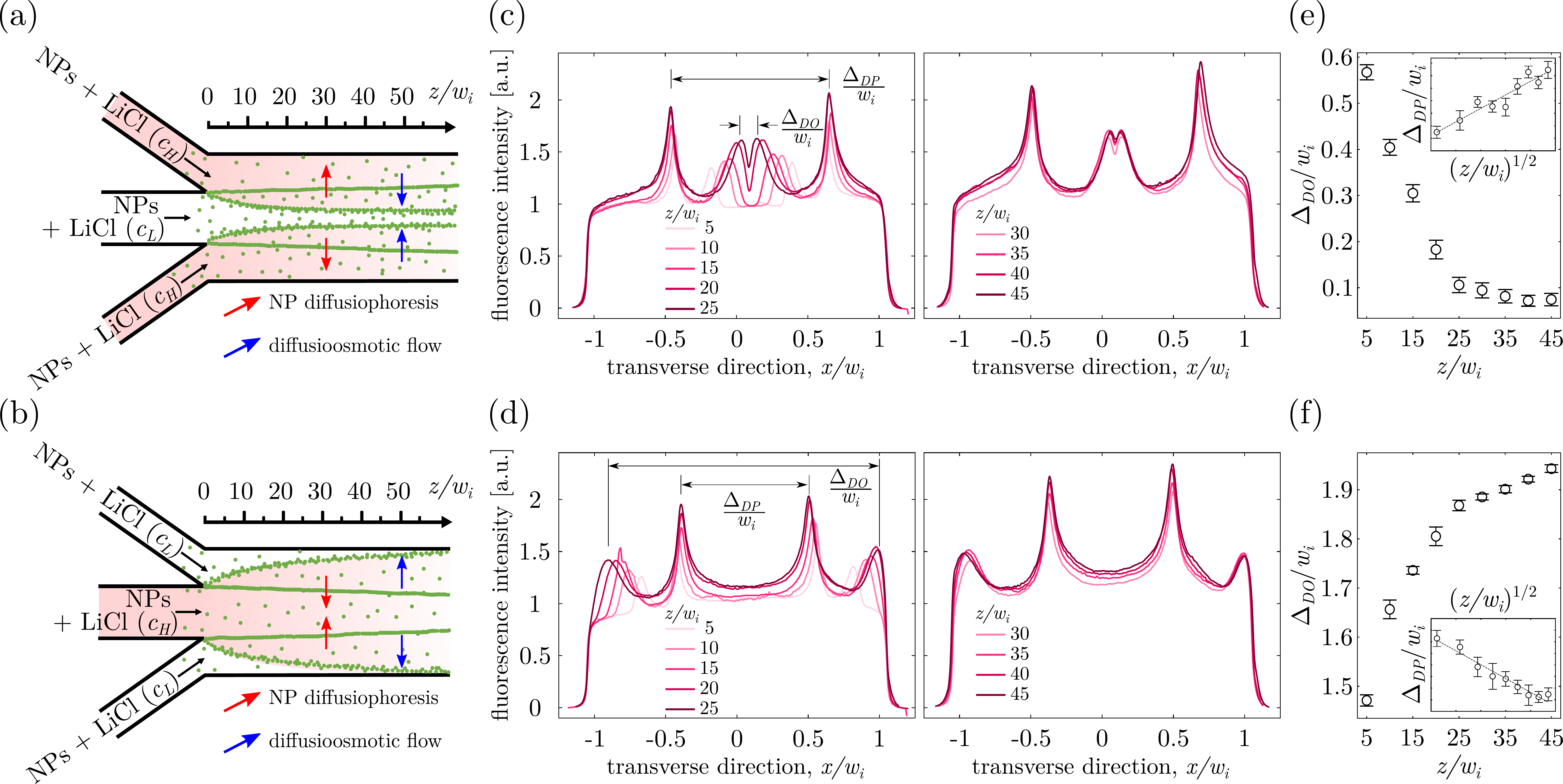} 
\end{center}
\caption{Relationship between the direction of the salt gradient and the
dynamics of the peak displacement  mechanism due to diffusioosmosis. (a,b) Schematic
diagrams of a single $\Psi$-junction microchip, where carboxylated NPs are
uniformly injected (in all channels) with two different configurations of salt
gradient. In (a), high concentration salt is in the outer channels ($\cH/c_{\sst L}$ = 100), whilst in (b) high salt concentration is in the inner channel.
Red-to-white shade qualitatively indicates the field of salt concentration.
Red (blue) arrows show the direction of diffusiophoresis transport (diffusioosmosis flow). 
(c,d) Normalized fluorescent intensity profile plots along the transverse direction
at various distances $z/w_i$ downstream the junction, for the configurations depicted in
schematics (a) and (b), respectively. (g,h) Normalized longitudinal distance $\DelDO$ plotted
with respect to the normalized distance $z/w_i$ for the
configurations (a) and (b), respectively.  \label{fig:DO_exp}} \end{figure}

The formation of peaks at the channel walls are in agreement with the observations reported in our previous study,
where we investigated the diffusiophoresis manipulation of charged nanoparticles in a $\Psi$-junction channel fitted with a microgrooved substrate~\cite{singh2020reversible}.
Our analysis revealed that the particle accumulation regions at the flat wall of the device were induced
by the vertical component of the salt concentration gradient which is originated by the Poiseuille-like velocity profile
in the rectangular microchannel~\cite{singh2020reversible}. 
Conversely, the convergence of the peaks in the $(z,x)$-plane towards lower
salinity regions is unprecedented. This particle behavior is also entirely unexpected as it seems to contradict the well-established interpretation
of particle diffusiophoresis~\cite{anderson1989colloid}, according to which colloids migrate towards higher
salinity regions with a diffusiophoresis velocity $\vecdp = \Gammadp \Nab ( \ln c)$
when the diffusiophoresis mobility $\Gammadp$ is positive – as it is the case for negatively charged particles
in LiCl solutions~\cite{Abecassis2008d}.
This implies that additional phenomena, such as
hydrodynamics and diffusioosmosis induced flows, must be involved in the
mechanism responsible for the transverse deviation of the peaks. 

To verify our hypothesis, two additional experiments with different
flow configurations were conducted (Fig.\,\ref{fig:DO_exp}a,b). First, a salinity gradient
was imposed as done in the experiment depicted above (high salt concentration
$\cH$ in the outer channels and low concentration $\cL$ in the inner channel).
However, contrary to what was done in the initial experiment, a homogeneous
solution of nanoparticles was injected in both inner and outer channels
(Fig.\,\ref{fig:DO_exp}a). In a second configuration, the chemical gradient was
swapped (high salt concentration in the inner channel,
Fig.\,\ref{fig:DO_exp}b). As showed by the fluorescence intensity
profiles along the transverse direction (Fig.\,\ref{fig:DO_exp}c,d),
the focused peaks induced by diffusiophoresis effects
were observed to converge ($\DelDO$ decreases with the distance) in the
first configuration (Fig.\,\ref{fig:DO_exp}a) and diverge
($\DelDO$ increases with the distance) in the second configuration
(Fig.\ref{fig:DO_exp}b). As
observed in the previous experiment, the variation of $\DelDO$ with $z$ (Fig.\,\ref{fig:DO_exp}d,e)
was rapid at shorter distances
from the junction ($z/w_i\lesssim25$) and more slowly at larger distances ($z/w_i\gtrsim25$).
By swapping the salinity gradient, we effectively interchanged the chemically
generated electrical field that gave birth to the fluid flow near the charged walls
induced by the diffusioosmosis slip velocity, $\vecdo = - \Gammado \Nab ( \ln c)$,
where the diffusioosmosis coefficient, $\Gammado$, has same sign and order of
magnitude of $\Gammadp$.
No physical parameters or properties in
our system were varied otherwise. 
Consequently, the systematic counter-intuitive transversal behavior of focused peaks in both
configurations (Figs.\,\ref{fig:DO_exp}d,e) suggests that they were deeply
affected by diffusioosmosis flows – which also justifies our choice of notation, $\DelDO$, for the peak distance. 
In other words, the peak displacement was always against the
diffusiophoresis-driven motion predicted for $\Gammadp > 0$ (red arrows in Fig.\ref{fig:DO_exp}a,b),
and, instead, it had always the same direction of the diffusioosmosis slip velocity
for $\Gammado>0$ (red arrows in Fig.\ref{fig:DO_exp}a,b). 
Note that in these two complementary experiments the presence of particles in both
inner and outer streams resulted in the formation of two additional, but narrower and more intense peaks, 
separated from each other by a distance $\DelDP$ (Fig.\,\ref{fig:DO_exp}c,d)
which evolved in an opposite way compared to $\DelDO$ (insets in Fig.\,\ref{fig:DO_exp}e,f). 
These peaks had been previously reported~\cite{Abecassis2009b} and are intrinsically different from
the ones discussed so far. Indeed, they form as the transverse $x$-component of the chemical gradient induces a
particle migration towards higher salt regions \emph{via} diffusiophoresis
only -- thus the notation $\DelDP$. This causes nanoparticles to accumulate along the entire depth of the channel and not just
nearby the channel walls~\cite{Abecassis2009b}.
In addition, the transversal migration of these peaks is extremely slow as their separation
distance $\DelDP$ is proportional to $z^{1/2}$, which is notably the
same scaling as $\Delta w$.
Conversely, the separation distance $\DelDO$ changes much faster with the
distance from the junction (Fig.\,\ref{fig:DO_exp}e,f). 

A numerical analysis was performed in COMSOL Multiphysics to confirm our
interpretation of the experimental observations.  The numerical hydrodynamic
velocity field $\vecu$, salt concentration $c$, and particle concentration $n$,
were calculated in a 3D domain consisting of a straight rectangular channel.  A
slip velocity  $\vecdo = - \Gammado \Nab ( \ln c)$ was imposed at the channel
walls and the particle velocity $\vecp$ was calculated as the sum of the hydrodynamic
velocity and the diffusiophoresis velocity, $\vecp=\vecu+\vecdp$.
The value of $\Gammado$ for the channel walls could not be measured, so it was used
as an adjusting parameter in the model to achieve a good match between experimental and numerical
results (see Numerical Methods section for details).
The simulated transverse profiles of the normalized particle concentration $n/n_0$, 
at increasing distances $z$ downstream the junction,
are shown in~Fig.\,\ref{fig:Fig1}d, and they are in good agreement with the fluorescence intensity
profiles measured in the experiments (Fig.\,\ref{fig:Fig1}b). It is worth noting that a close quantitative match
is not expected since the experimental profiles correspond to the convolution
of the particle fluorescence intensity with the microscope point-spread function~\cite{gu2000advanced},
whereas the numerical profiles are directly obtained from the simulated particle concentration field by
averaging the concentration over the channel depth ($y$ direction).
A good quantitative agreement between experiments and simulations can be seen for the peak separation
$\DelDO$ at increasing distances from the junction (Fig.\,\ref{fig:Fig1}c), which is consistent
with the fact that the effect of the microscope point-spread function on $\DelDO$ measurements should be negligible.
Fig.\,\ref{fig:Fig1}e shows the simulated particle concentration field on the plane perpendicular to
the flow direction at $z/w_i=25$, and this is in a good agreement with the confocal image of the channel
cross-section at the same distance from the junction (Fig.\,\ref{fig:Fig1}a).
As expected, 
the salt concentration isolines 
in the inner region of the channel ($|x|/w_i<0.5$),
showed in Fig.\,\ref{fig:Fig1}e,
are bent towards the outer flow 
 because of the Poiseuille-like hydrodynamic velocity profile. Consequently, the onset of a vertical
component of the salt concentration gradient and, thus, of the diffusiophoresis velocity, causes the accumulation
of particles at the top and bottom walls. 
The diffusioosmosis slip velocity at the walls induces the formation of four symmetric recirculation regions
in the in-plane total particle velocity field $\vecp$, whose streamlines are showed by white arrows in Fig.\,\ref{fig:Fig1}e. 
As a result, the accumulated particles are advected 
along the top and bottom walls toward the central region of the channel, namely from higher to lower
salt concentration regions. To summarize, the observed particle behavior is governed by
a combination of particle diffusiophoresis along the vertical axis, which induces particle accumulation, 
and diffusioosmosis flow along the horizontal axis, which pushes the accumulation peak towards the center of the channel. 

\subsection{Particle size detection and size-based separation in double-junction devices}

\begin{figure}[H]
\begin{center}
\includegraphics[width=\columnwidth]{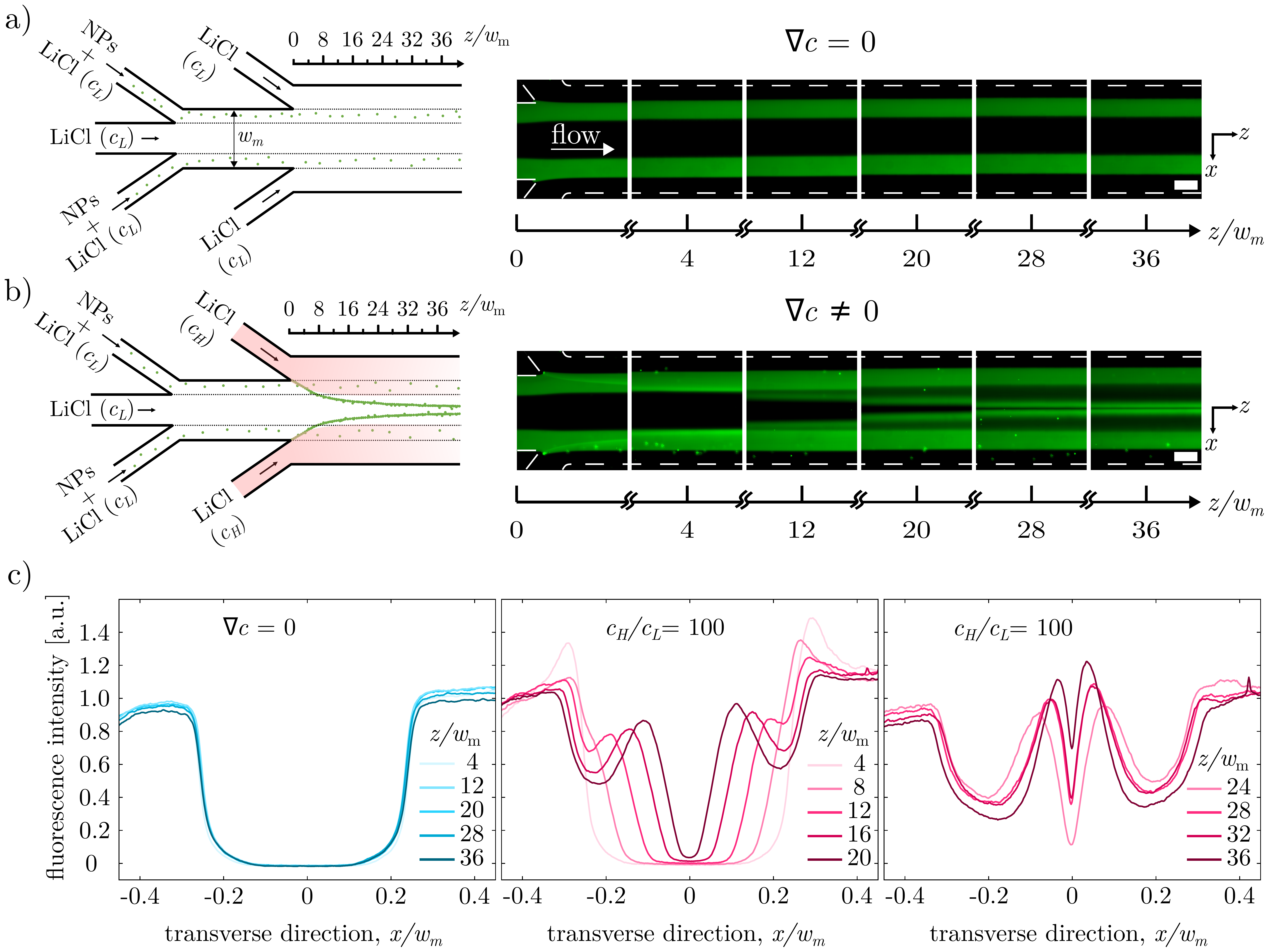}
\end{center}
\caption{Particle focusing and fractioning in a double $\Psi$-junction device.
(a,b) Schematic diagrams of the double-junction microchip and the corresponding
epi-fluorescence micrographs showing the fluorescent
carboxylate polystyrene nanoparticles ($\textit{d}$ = 549.8 $\pm$6.8 nm, $\zeta$=
-50.4 $\pm$0.2 mV) at various distances from the junction when there is no salinity
gradient ($\nabla$\textit{c} = 0) in (a) and when there is a salinity gradient
($c_{H}$/$c_{L}$=100) in (b). 
The micrographs were acquired by positioning the focal plane nearby the bottom wall of the device.
Scale bar is 75 $\upmu$m. (c) Normalized
fluorescent intensity profiles when there is no salinity gradient (blue
curve) and  when there is a salinity gradient (red curve) at various distance
downstream the junction. \label{fig:2JD}} \end{figure}

The $\Psi$-junction microchip, depicted in Fig.\,\ref{fig:Fig1}a,
could be potentially adopted for the online pre-concentration, via
solute-driven transport, of charged synthetic or biological colloids,
including macromolecules~\cite{lechlitner2018macromolecule},
liposomes~\cite{shin2016size}, exosomes~\cite{rasmussen2020size},
viruses~\cite{Palacci2010b} and bacterial cells~\cite{shim2021co}. This can be
particularly useful for applications such as microfluidic point-of-care
diagnostics and point-of-need bioanalytical testing, provided that the target 
analytes are charged and, thus, susceptible to diffusiophoresis migration. 
Alternatively, the same device could be used for the solute-driven
accumulation of charged nanoparticles that are conjugated to recognition moieties, such as
antibodies or aptamers, for the capture and detection of the target molecules~\cite{azzazy2009vitro}. 
Furthermore, since the diffusiophoresis mobility depends on both particle
size and zeta potential~\cite{prieve1984motion,shin2016size}, the
diffusiophoresis-driven accumulation of nanoparticles at the device walls could be
exploited also for particle characterization, fractioning (commonly known as field flow fractioning) and sorting.
However, the device configuration
showed in Fig.\,\ref{fig:Fig1}, does not lend itself to such
applications, because the accumulation peaks are advected towards the central region of the channel, thus
overlapping with the bulk colloidal stream. Consequently, particle fractioning and
sorting are not possible. Moreover, the fluorescence intensity
of the accumulation peaks is partially screened by the background fluorescence signal
generated from the colloids in the bulk (see Fig.\ref{fig:Fig1}a,b), thus limiting the accuracy of the peak intensity detection and
hampering the ability to characterize particles by charge or size.
These limitations could be overcome if the accumulation peaks migrated away from the bulk colloidal stream.
To achieve this, first we tested a different flow configuration, whereby a salinity gradient was imposed
as done in the experiment of Fig.\,\ref{fig:Fig1} –- namely higher salt concentration $\cH$ in
the outer channels and lower salt concentration $\cL$ in the inner channel -- but the
colloidal particles were present in the outer stream only. However, this test came
to no avail (see Fig.\,S2 in Support Information Material), since the accumulation peaks did not form.
This is because  the salt concentration isolines, shown in Fig.\,\ref{fig:Fig1}e, are bent outwards
only within the inner stream region of the channel ($|x|/w_i\le0.5$), but
no colloids are now present in that region. 
As a result, the vertical component of the salt gradient
no longer leads to the migration of colloids towards the top and bottom
walls and the consequent formation of the particle accumulation peaks.
Therefore, a new chip design was required to exploit the observed phenomenon of
particle focusing for particle fractioning, separation and characterization.  
Fig.\,\ref{fig:2JD}a,b depict the blueprint of the double $\Psi$-junction microchip designed for this purpose.
The channel geometry is similar to the one of the previous device, but 
a second $\Psi-$junction was added upstream of the first one. 
The downstream junction is used to regulate the salt concentration gradient in the device by swapping
the outer flow stream between a low salt concentration ($\cL$) solution, leading to no salt gradient (Fig.\,\ref{fig:2JD}a),
and a high salt concentration ($\cH$) solution, generating a steady-state salt gradient (Fig.\,\ref{fig:2JD}b). 
The upstream junction allows to control the position of the colloidal stream within the main channel.
All streams injected from the inlet channels of the upstream junction have a low salt concentration ($\cL$), but only
the outer streams are laden with colloidal particles. Consequently, the inner region of the channel
remains particle-free so that, upon imposition of the salt gradient, the focused particle peaks
can converge into this region without overlapping with the bulk colloidal stream.

This new double-junction device was first tested using carboxylate
polystyrene nanoparticles ($\textit{d}$ = 549.8 $\pm$6.8 nm, $\zeta$=
-50.4 $\pm$0.2 mV). Epifluorescence micrographs were taken at different
distances downstream the junction, $z/w_m$, either in absence or presence of a
salt concentration gradient (Fig.\,\ref{fig:2JD}a,b).  Note that for the
double-junction device, the width of the inner channel of the downstream
junction, $w_m=250\,\upmu$m, is used as the characteristic channel size and
all lengths are normalized with respect to it.  The corresponding fluorescent
intensity profiles along the transverse direction $x$ show that peak formation
occurred when a salinity gradient was imposed (red curves in
Fig.\,\ref{fig:2JD}c). Crucially, the peaks moved toward the inner region of
the channel, which would otherwise be empty in the absence of a salt
concentration gradient (blue curves Fig.\,\ref{fig:2JD}c).
\begin{figure}[H]
\begin{center}
\includegraphics[width=\columnwidth]{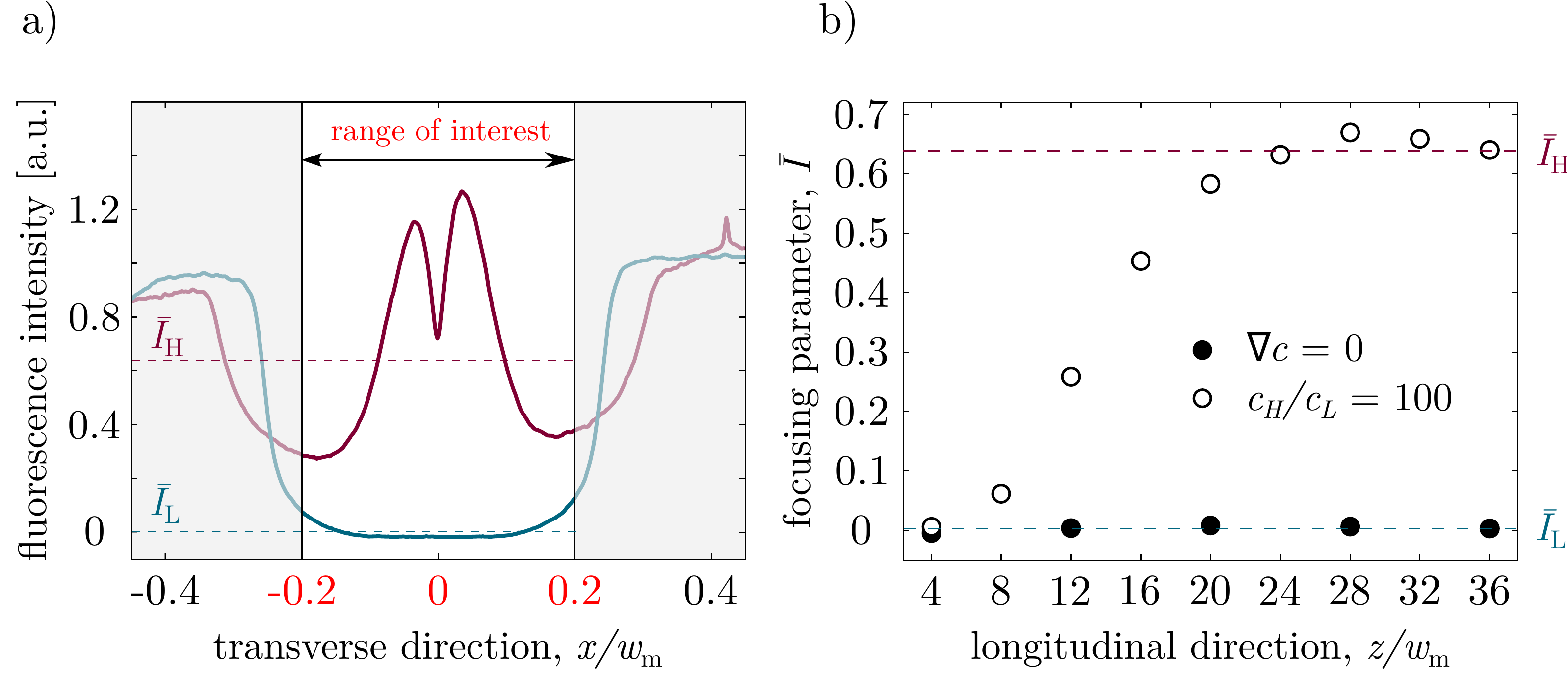}
\end{center}
\caption{Focusing parameter in double-junction device for carboxylate polystyrene nanoparticles 
($\textit{d}$ = 549.8 $\pm$6.8 nm, $\zeta$=
-50.4 $\pm$0.2 mV)
(a) Normalized fluorescent
intensity profile at $z/w_m=36$
without salinity gradient (blue curve) and with salinity gradient
(red curve). The region of interest corresponds to a selected central
region of the channel, $x/w_m\in[-0.2,0.2]$. The focusing parameter, $\bar{I}$, is calculated as the
average fluorescence intensity over the region of interest. (b) Focusing parameter 
at different distances downstream the junction ($z/w_m$) under no salinity gradients (solid circles) and
with a salinity gradient (empty circles). The blue and red dashed lines are the
focusing parameter values corresponding to their respective intensity plots in
(a) at $z/w_m$ = 36.  \label{fig:f_param}} \end{figure}
To quantify this focusing effect, we introduce the focusing parameter,
$\bar{I}$, defined as the average value of the normalized 
fluorescent intensity profile within the range of interest, $x/w_m \in
[-0.2,0.2]$, which is equivalent to a transverse section, ca. $50\,\upmu$m
wide, located at the center of the channel (Fig.\,\ref{fig:f_param}a).  Note
that the width of the region of interest is chosen to exclude the fluorescence
intensity generated by the particles in the bulk colloidal stream, namely the
gray shaded regions in Fig.\,\ref{fig:f_param}a.  Fig.\,\ref{fig:f_param}b
shows the focusing parameter at increasing distances $z/w_m$ downstream the
junction in the presence (empty circles) and absence (solid circles) of a salt
concentration gradient. Under the examined conditions, at $z/w_m=36$, the
focusing parameters were $\bIH=0.66$ and $\bIL=0.03$ with and without a salt
concentration gradient, respectively.

\begin{figure}[t!]
\begin{center}
\resizebox{6.5in}{!}{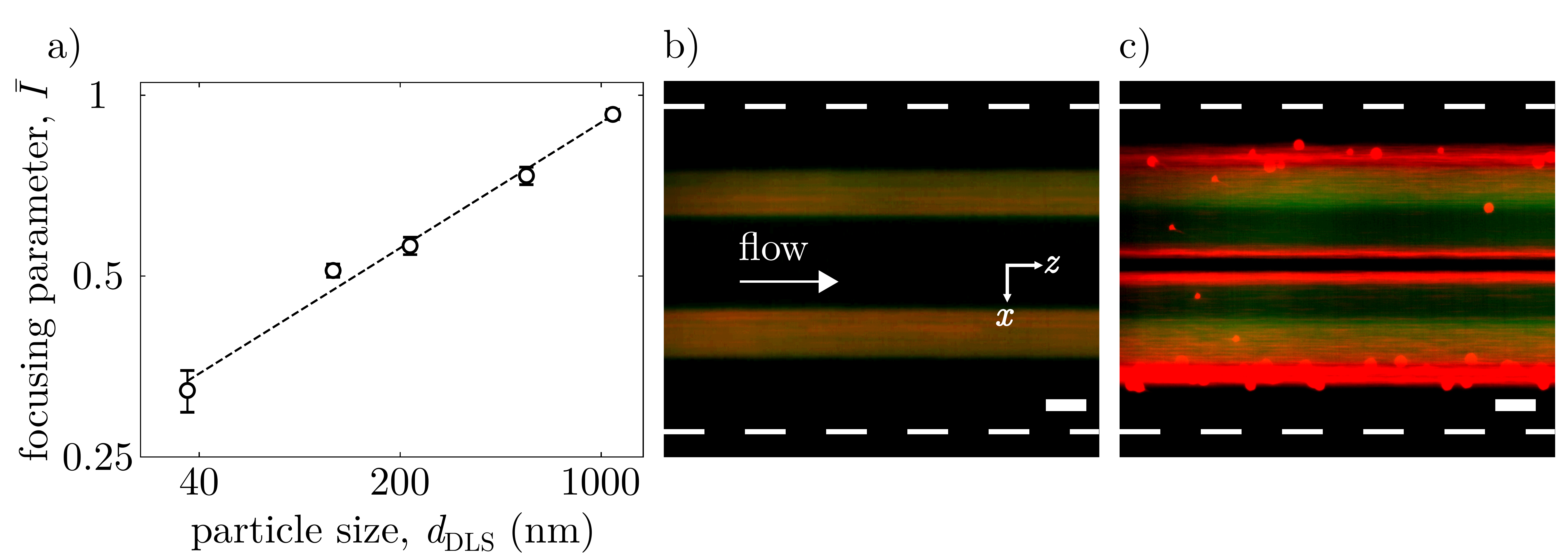}
\end{center}
\caption{Particle size characterization and particle separation. (a) 
Average focusing parameter for different particle sizes at approximately the
same zeta potential (see Tab.\,\ref{tab:I_d}).
(b-c) Epi-fluorescence images taken at
$z/w_m$ = 36 showing particle dynamics of mixed yellow green fluorescent
(505/515) 36 nm  and red fluorescent (580/605) 1098 nm carboxylate polystyrene colloids under
no salinity gradient (b) and with  salinity gradient (c). Scale bar is
$50\,\upmu$m. White dashed line corresponds to
the microfluidic channel boundaries. \label{fig:size}} \end{figure}

Since the formation of the accumulation peaks is driven by the diffusiophoresis
migration of charged nanoparticles from the channel bulk toward the top
and bottom walls of the microchannels, it
is reasonable to expect that the peak intensity, and thus the focusing
parameter, can be correlated to the diffusiophoresis coefficient of the
nanoparticles.  For a charged nanoparticle in an electrolyte solution, the
diffusiophoresis coefficient $\Gammadp$ is an increasing function of the
particle size~\cite{prieve1984motion,shin2016size}, if the thickness of the
Debye layer $\kappa^{-1}$ formed around the particle is a few percent greater
than the particles radius $a$, namely $(\kappa a)^{-1}\gtrsim 1\%$. On the
other hand, for $(\kappa a)^{-1}\rightarrow 0$ the coefficient, $\Gammadp$,
levels off to a constant value that is independent of the particle size.  In a
0.1~mM LiCl solution, the Debye layer thickness is $\kappa^{-1}=32\,$nm,
therefore it is expected that for sub-micron particles ($2a\lesssim
1\,\upmu$m and $(\kappa a)^{-1}> 6\%$ ), the coefficient $\Gammadp$ and the accumulation peak intensity
increase with the particle size.  To verify this hypothesis, we measured the
focusing parameter for carboxylate polystyrene particles with a diameter
ranging from tens nanometers to one micrometer. Note that the zeta potential of the
particles were very similar for all diameters (Tab.\ref{tab:I_d}).  From the
logarithmic plot in Fig.\,\ref{fig:size}a, it is apparent that the focusing
parameter $\bar{I}$ increases with the particle diameter, $\dDLS$, determined
via dynamic light scattering (DLS).  Interestingly, $\log \bar{I}$ and $\log
\dDLS$  are linearly correlated  ($R^2=0.99$).
\begin{equation} \log \bar{I} = 0.299\, \log \dDLS -0.941 \label{eq:I_d}
\end{equation}
where the value of $\dDLS$ is expressed in nanometers.  Consequently,
Eq.\eqref{eq:I_d}  can be effectively used as a calibration line for the
microfluidic characterization of the size of nanoparticles of similar zeta potential, whereby the focusing
parameter $\bar{I}$ is measured experimentally  to determine the nanoparticle
diameter
 \begin{equation} \dMF = \left({8.73}~{\bar{I}}
\right)^{3.34} \label{eq:d_I} \end{equation}
with the value of the microfluidically measured diameter, $\dMF$, expressed in nanometers.  The uncertainty
$\sigma_d$ on $\dMF$ can be calculated from the experimental uncertainty
$\sigma_I$ on $\bar{I}$ via error propagation in Eq.\eqref{eq:d_I},
which leads to $\sigma_d^2 = 11.18\, \dMF^2 \sigma_I^2/\bar{I}^2$.
Tab.\,\ref{tab:I_d} shows a comparison between the particle diameter, $\dDLS$,
measured via DLS, and the particle diameter, $\dMF$, measured via the proposed
microfluidic method.  From the absolute relative difference between the DLS and
microfluidic measurements, $\varepsilon_r=|\dDLS-\dMF|/\dDLS$, it can be
concluded that our microfluidic device can be successfully used to measure the
diameter of sub-micron particles within a reasonable accuracy.

\begin{table*}[t!] 
\caption{
Comparison between dynamic light scattering (DLS) and microfluidic (MF) characterization
of sub-micron particle size. $\bar{I}$ measured focusing parameter, $\dDLS$ particle size
determined via DLS, $\zeta$ particle zeta potential, $\dMF$ particle size
determined microfluidically via Eq.\eqref{eq:d_I}, $\varepsilon_r$ absolute relative difference
between $\dDLS$ and $\dMF$. The error for $\dDLS$ and $\zeta$ represent the standard deviation obtained from the instrument. The error for $\bar{I}$ is the standard error from experimental mean values.}

\begin{center} \begin{tabular}{C{0.15\textwidth}
C{0.15\textwidth} C{0.15\textwidth} C{0.15\textwidth} C{0.15\textwidth} } 
\toprule \textbf{$\bar{I}$} &
\textbf{$\dDLS$} (nm) & $\zeta$ (mV)  & \textbf{$\dMF$} (nm) & $\varepsilon_r$ \\
\hline
0.323 $\pm$ 0.03 & 36 $\pm$ 0.2 & -58.3 $\pm$ 1.4 &32 $\pm$ 9 & 12\%   \\
0.511 $\pm$ 0.01  & 117 $\pm$ 0.5 & -57.6 $\pm$ 1.2   &150 $\pm$ 13  & 28\%   \\
0.561 $\pm$ 0.02  & 217 $\pm$ 0.9  & -54.9 $\pm$ 0.7   &206 $\pm$ 23 & 5\%   \\
0.735 $\pm$ 0.03  & 550 $\pm$ 6.8 & -50.4 $\pm$ 0.2    &505 $\pm$ 57 & 8\%   \\
0.929 $\pm$ 0.02 & 1098 $\pm$ 29 & -55.5 $\pm$ 0.7   &1108 $\pm$ {70} & 1\%   \\
\bottomrule \end{tabular} \end{center} 
\label{tab:I_d} \end{table*}

The combined effects of peak formation and drift towards the center of the channel can be exploited also for 
particle size-based separation.
To demonstrate this additional application of the double-junction chip,
a binary colloidal mixture of 36~nm and $1.098\,\upmu$m diameter particles was injected into the device.
The carboxylate polystyrene particles were stained with fluorophores with different
emission peaks, namely
515~nm (yellow-green) for the smaller particles and 605~nm (red)
for the larger particles. In absence of a salt concentration gradient,
both populations of particles were advected by the flow and remained confined
in the same regions of the channel where they were initially
injected (Fig.\,\ref{fig:size}b). Upon imposition of a salt contrast
($\cL=0.1\,$mM, $\cH=10\,$mM), the larger (red) particles 
were strongly focused at the center of the channel where they formed two intense peaks. Conversely,
the smaller (yellow-green) particles did not form any accumulation peak and only a small fraction of them
drifted toward the center.
Furthermore, the two colloidal bands made of larger particles, expanded 
along the transverse ($x$) direction
towards the higher salt concentration (outer) regions of the channel due to diffusiophoresis~\cite{Abecassis2008d},
with two additional focusing peaks forming at these locations.
On the other hand, the colloidal bands made of smaller particles did not drift
towards the outer regions because of their smaller
diffusiophoresis coefficient.  As a result, diffusiophoresis and diffusioosmosis
could be effectively exploited for the microfluidic separation and sorting of the two populations of particles according to their size.

\subsection{Detection of liposome zeta potential and membrane composition in double-junction devices}

The range of applications of the double-junction device can be expanded further
by leveraging the dependence of the diffusiophoresis coefficient on zeta
potential, which is a key property of colloidal system.
To this end, we used nano-sized unilamellar liposomes whose
 surface charge and, thus zeta potential, can be easily
tuned by adjusting the lipid membrane composition. Such nanoparticles,
which are often used as drug carriers~\cite{sercombe2015advances,guimaraes2021design}
and cell membrane models~\cite{el2008liposomes}, are hence
suitable for investigating the effect
of zeta potential on the focusing phenomenon in the double-junction device.

\begin{figure}[tb!]
\begin{center}
\includegraphics[width=\columnwidth]{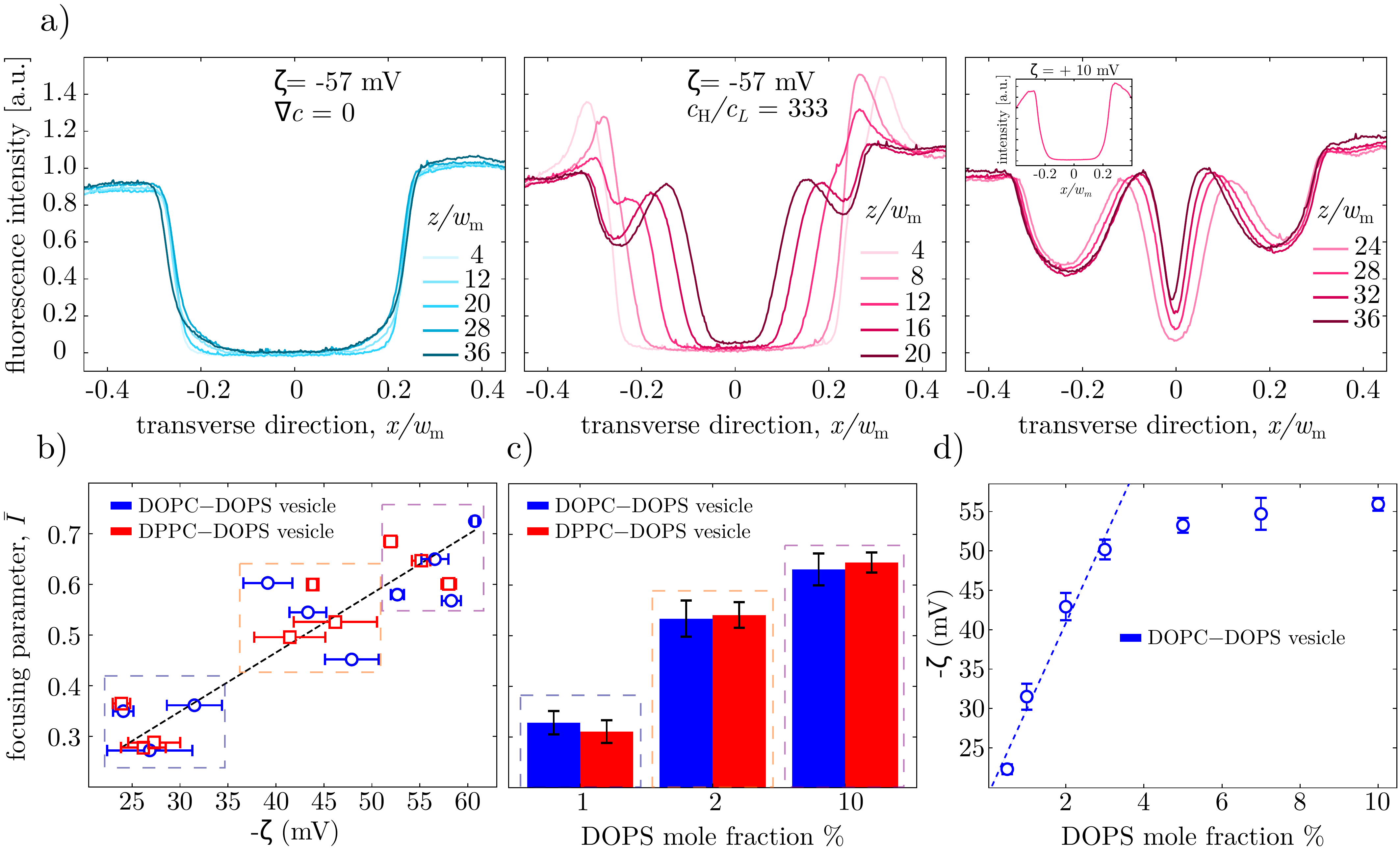}
\end{center}
\caption{(a) Normalized fluorescent intensity profiles along the transverse
direction $x$ and at varying distances $z$ from the downstream junction without
(blue curves) and with (red curves) a salinity gradient for negatively
charged ($\zeta=-57$~mV) 10:90 DOPS:DPPC liposomes. 
Inset shows the fluorescent intensity profile for zwitterionic DPPC liposomes
($\zeta=+10$~mV).
(b) Experimental
focusing parameter against zeta potential for DOPC (blue) and DPPC (red) based
liposomes with varying anionic DOPS concentrations (navy box = 1 PS, orange = 2
PS, purple = 10 PS . (c) Experimental focusing parameter against the anionic DOPS
lipid content. (d) Relation between vesicle zeta potential and anionic DOPS
lipid content for DOPC-DOPS vesicles. The dashed line corresponds to the
linear regression of the first four data points.  
Fabricated liposomes had an average size of $\textit{d}$
= 166 $\pm$ 4 nm.  \label{fig:liposomes}} \end{figure}

In the first set of experiments, we fabricated negatively
charged liposomes by adding 10 $\%$ mol fraction of the anionic phospholipid
1,2-dioleoyl-sn-glycero-3-phospho-L-serine (DOPS)
to the zwitterionic DPPC phospholipid ($\textit{d}$ = 166 $\pm$
4 nm, $\zeta$ $\approx$ -57 mV, polydispersity index $\approx$ 0.1). The phophoserine
headgroup (PS) was specifically chosen as its presence in the outer leaflet of the membranes
of biological particles can be associated with pathological and physiological processes,
such as in tumor-derived exosomes ~\cite{maia2018exosome} and during cell apoptosis~\cite{fadeel2009ins}.
The vesicles were dispersed in a low salt concentration ($\cL$ = 0.01 mM) 
LiCl solution buffered with HEPES salt and EDTA chelating agent (pH = 8.15).
The aqueous stream of charged liposomes was pumped in
the outer channels of the upstream junction, whereas the same
low salt concentration solution without liposomes was injected in the
inner channel of the device. Finally, a high salt concentration ($\cH$ = 6.65 mM) LiCl solution, also buffered
with HEPES salt and EDTA, was pumped in the outer channels of the downstream junction.
The fluorescence intensity profiles along the transverse direction are showed in Fig.~\,\ref{fig:liposomes}a.
It can be observed that charged liposomes display a similar behavior to the one observed for
polystyrene nanoparticles.
Upon imposition of a salt gradient (red curves), peaks of accumulated liposomes formed
and converged to the center of the device ($|x|/w_m\le0.2$) and away from the colloidal bulk streams ($|x|/w_m>0.2$).
Conversely, when the salt concentration was the same throughout the
device (blue curves), liposome peaks did no form  and no particles migrated toward the inner region of the channel.
Note that for liposomes, a higher salt contrast ($\cH/\cL = 333$) was adopted
in comparison to the one applied in the experiments with carboxylated
polystyrene nanoparticles ($\cH/\cL = 100$).  Indeed, in agreement with
previous studies~\cite{shin2019osmotic,rasmussen2020size}, the migration speed
of liposomes under salt gradients is typically smaller that the one of
polystyrene nanoparticles with comparable size and zeta potential, therefore higher salt
contrasts are required. This is likely due to the soft and water-permeable
nature of the liposomes that may trigger additional migration mechanisms, such
as osmophoresis~\cite{anderson1986transport}, or affect the diffusiophoresis
response through membrane
permeability~\cite{anderson1983movement,yeh2018diffusiophoresis}, membrane
viscosity and vesicle shape deformations.  To confirm that the imposed salt
contrast did not affect the liposome stability and size in our experiments, a
liposome solution in low salt concentration ($\cL$) buffer was analyzed via
dynamic light scattering before and after dilution in high salt concentration
($\cH$) buffer, resulting in no detectable changes in the particle size
distribution.  Interestingly, the inset in Fig.\,\ref{fig:liposomes}a shows
that when using zwitterionic DPPC liposomes ($\zeta=+10\,$mV) instead of
negatively charged DPPC-DOPS liposomes, no focusing effect is observed under a
salinity gradient. This observation is consistent with our physical
interpretation of the particle focusing phenomenon. Indeed, DPPC liposomes in
the buffer solution carry a very weak positive charge and, therefore, do not
migrate by diffusiophoresis towards the top and bottom non-zero charged walls
of the channel.  In absence of a particle accumulation nearby the walls, the
diffusioosmosis flows at the walls do not affect the colloid distribution in
the device and no particles are directed toward the center of the channel
($|x|/w_m\le 0.2$).  Remarkably, this finding could be exploited for the
microfluidic separation of liposomes based on their surface charge. 
Indeed, for a mixture of negatively charged liposomes and zwitterionic lipid vesicles,
only the charged liposomes will migrate towards the central region of the microchannel,
whereas the trajectories of the zwitterionic lipid vesicles won't be affected by the salt concentration
gradient and they will keep clear of the central region of the channel.

In another set of experiments, we investigated the effect of the lipid
composition of the liposome membrane on  particle focusing.  Specifically, we
adjusted the zeta potential of the liposomes by varying the amount of charged
lipid content in the membrane composition, and we controlled the
viscosity/fluidity of the membrane by using lipid mixtures with a fluid/gel
phase transition temperature either above or below the room temperature. Since
altering the zeta potential of the liposomes affect significantly their
diffusiophoresis mobility, it should be possible to correlate the charged lipid
content with the intensity of the focusing effect in the device. On the other
hand, it is known that the diffusiophoresis speed depends also on the Newtonian
and Maxwell stress balance at the particle
surface~\cite{yang2018diffusiophoresis}, which in turn depends on 
the viscosity of the membrane. However, to date the role of membrane
viscosity on the diffusiophoresis of lipid vesicles or living cells has yet to be explored.
To carry out this analysis, we considered six
different liposome populations at 1,2 and 10$\%$ DOPS (charged lipid) mole fraction added to 
zwitterionic lipids, either 1,2-dioleoyl-sn-glycero-3-phosphocholine (DOPC) 
for a disordered fluid phase membrane or 1,2-dipalmitoyl-sn-glycero-3-phosphocholine (DPPC) 
for an ordered gel phase membrane.
Three independent samples were produced for each liposome population, and for each sample both
the zeta potential and the corresponding focusing parameter were determined. All liposome populations had 
similar size ($\textit{d}$ = 166 $\pm$ 4 nm) and polydispersity index.
Fig.\,\ref{fig:liposomes}b and Fig.\,\ref{fig:liposomes}c show how the focusing parameter $I$ is positively correlated with
the particle zeta potential and the DOPS (charged lipid) molar fraction. Remarkably, for a given DOPS concentration,  there is no
statistically significant difference between the focusing parameters observed for DOPC-DOPS fluid-like membrane vesicles (blue bars/symbols)
and for DPPC-DOPS gel-like membrane vesicles (red bars/symbols), thereby showing the irrelevance of the liposome membrane viscosity
for the diffusiophoresis transport under the examined conditions. 
The linear regression (dashed line) between liposome zeta potential and focusing parameter is given by 
\begin{equation}
\bar{I}=-0.0116\,\zeta
\label{eq:I_zeta} 
\end{equation}
with $\zeta$ expressed in millivolt.
Eq.\eqref{eq:I_zeta} can be used as a calibration curve for the microfluidic detection of the zeta potential of similarly-sized
particles, whereby the focusing parameter $\bar{I}$ is measured experimentally to quantify the zeta potential according to the following relation
\begin{equation}
\zeta=-86.2\,\bar{I}
\label{eq:zeta_I} 
\end{equation}
Furthermore, we conducted an electrophoretic light scattering analysis to correlate
the zeta potential of DOPC-DOPS vesicles with the DOPS (anionic lipid) content in the membrane (Fig.\,\ref{fig:liposomes}d).
As the DOPS molar fraction increases, the magnitude of the zeta potential of liposomes increases, and 
eventually it plateaus at ca. 8-10$\%$ DOPS content. Beyond this point, adding more DOPS
to the membrane does not affect the zeta potential, due to the formation of a charged condensed layer of Na$^+$
counterions around the outer leaflet of the liposomes~\cite{lutgebaucks2017characterization}. 
On the other hand, the zeta potential is highly sensitive to the DOPS content
at low anionic lipid concentrations ($\le\%3$). This suggests a strategy for the application of the double-junction device
for the quantification of small amounts of DOPS lipids in the outer leaflet membranes of liposomes. 
Indeed, for low anionic lipid concentrations ($\le\%3$), the relation between zeta potential $\zeta$ (expressed in millivolt) 
and the DOPS molar fraction $\xDOPS$ (expressed in percentage values) 
can be well approximated by the following linear relationship, 
plotted as a dashed line in Fig.\,\ref{fig:liposomes}d,
\begin{equation}
\zeta=-10.9 \, \xDOPS-19.0
\label{eq:zeta_DOPS}
\end{equation}
By combining Eq.~\eqref{eq:zeta_DOPS} and Eq.~\eqref{eq:zeta_I}, it follows
\begin{equation}
\xDOPS = 7.91\, \bar{I} - 1.74
\label{eq:DOPS_I}
\end{equation}
which allows one to estimate the DOPS lipid molar fraction from the experimental measurements of the focusing parameter $\bar{I}$.
To conclude,  Eq.~\eqref{eq:zeta_I} and Eq.~\eqref{eq:DOPS_I} show how
the double-junction device can be used effectively for the quantitative estimate of both the zeta
potential and the DOPS lipid content of the outer leaflet of liposome membranes.


\section{CONCLUSIONS}
We described an unreported mechanism where diffusiophoresis and diffusioosmosis
are closely intertwined, leading to a strong transverse focusing of nanoparticles
in a single $\Psi$-junction microchannel under continuous and steady axial flow conditions. 
Parallel electrolyte streams are merged at the junction of the device to generate a chemical gradient in both the transverse and
vertical directions. As a result, the particles first migrate vertically by diffusiophoresis from the bulk towards the top and bottom walls,
and subsequently undergo a transverse horizontal migration along these walls
driven by diffusioosmosis. 
The remarkable coupling of diffusiophoresis and diffusioosmosis along two perpendicular directions
allows us to take advantage of the relatively weak diffusioosmotic slip velocities 
in a pressure-driven microfluidic flow without resorting to dead-end or nano-sized channels.
Indeed, here it is the vertical diffusiophoretic migration that confines the particles nearby the 
charged walls where the effects of diffusioosmotic flows are the most intense.
Consequently, one can avoid the most common drawbacks associated with dead-end pores or nano-confined channels,
such a risk of device obstruction and clogging, costly device fabrication procedures and difficult recovery of the 
colloidal sample.
By means of epi-fluorescence microscopy, two accumulation peaks are formed and their separation distance
decreases with the distance from the channel inlet, eventually reaching a
plateau value that depends on the size and $\zeta$-potential of the colloids.
Those peaks are of an intrinsically different nature of those
observed in previous works~\cite{Abecassis2008d}, the dynamics of which was
exclusively driven by diffusiophoresis. 

We also showcased the exploitation of this novel mechanism
for the continuous separation and characterization of colloidal particles.
A proof-of-concept double-junction device was developed and used for the accurate measurements of
the diameter of colloidal beads with same $\zeta$-potential, based on the measurement of an ad-hoc,
purposely defined, focusing parameter $\bar{I}$. We also demonstrated how the
dependence of the transverse drift dynamics of the accumulation peaks on the colloid size
can be further leveraged to separate large particles ($1.098\,\mu$m) from
smaller ones ($36\,$nm) in a bi-modal mixture. Moreover, the setup is proven
to allow for a reasonable assessment of the $\zeta$-potential of same size
particles through the measurement of the focusing parameter $\bar{I}$. Based
on this latter principle, we eventually showed how, in the case of liposomes, the
measurement of the $\zeta$-potential can be used to assess the
chemical composition of the membrane (here the molar fraction of the charged DOPS lipid),
 at least in the low concentration limit. 
Note that, whilst previous studies allowed only for batch measurements of particles located in dead-end pores or open nanochannels,
our double-junction device enables the online, continuous and high-throughput characterization
of particles directly within the colloid stream. This also facilitates the recollection
of the analyzed sample for further off-chip downstream analysis. 
Finally, we showed how, under the examined
experimental conditions, the fluid-like or gel-like states of the membrane, and thus the membrane viscosity,
does not affect the diffusiophoretic response of the liposomes.

We envisage that this study will open up new exciting routes for the use of
diffusiophoresis
and diffusioosmosis for the microfluidic manipulation and characterization of both synthetic and natural particles.
Our microfluidic devices can unlock a wide scope of possibilities related to bio-oriented applications,
such as the pre-concentration, sorting, sensing and analysis of biological entities, including
liposomes, extracellular vesicles and bacteria. By relying on the chemical energies of the electrolyte
solutions rather than on external energy sources
and by adopting cheap and easy-to-fabricate microfluidic chips, the proposed particle manipulation and characterization
strategies naturally lend themselves to the development of portable, cost-effective, point-of-need microdevices
for chemical and biochemical analysis, diagnostics, drug screening and drug delivery applications.

\section{METHODS}
\subsection{Materials}

Invitrogen\texttrademark~ FluoSpheres\texttrademark~ , carboxylate-modified nanoparticles, red fluorescent (580/605), 2 $\%$ solids, were purchased from ThermoFisher scientific at various sizes (20,100,200,500,1000 nm). In addition, Invitrogen\texttrademark~ FluoSpheres\texttrademark~ , yellow-green fluorescent (505/515), 2$\%$ solid, 20 nm carboxylate-modified nanoparticles was also purchased for separation experiments from the same supplier. Lithium chloride salt (LiCl, 99\%) used for diffusiophoresis experiments was purchased
from Acros Organics. Aqueous liposome solutions were prepared with buffer salt,
HEPES and chelating agent EDTA purchased from Sigma Aldrich. 
Lipophilic dye, 3,3'-dioctadecyloxacarbocyanine perchlorate (DiO$_{c_{18}}$) -- used
for staining liposomes -- and chloroform (99\%) -- used for preparing lipid films --
were purchased from Sigma Aldrich. RTV 615 polydimethyl siloxane used for the
fabrication of microfluidic devices was purchased from Techsil, UK. 
The phospholipids 1,2-dioleoyl-sn-glycero-3-phosphocholine (DOPC),
1,2-dipalmitoyl-sn-glycero-3-phosphocholine (DPPC) and 
1,2-dioleoyl-sn-glycero-3-phospho-L-serine
(DOPS) were purchased from Avanti Polar Lipids. 
Deionized water (18.2 M$\Omega$\,cm) produced from an ultrapure milli-Q grade
purification system (Millipore, USA)  was used to prepare all aqueous
solutions.

\subsection{Fabrication and operation of the microfluidic devices \label{seubsec:mufluidic_device}}

Standard photo- and soft-lithography techniques were employed in manufacturing
the microfluidic devices. Briefly, the CAD drawings of both one and two-junction 
devices were printed on a photomask film (Micro Lithography Services, UK) 
and subsequently used to produce an SU-8 master mold
of ca. 50 $\upmu$m thickness on a silicon wafer (Inseto, UK). Imprinted
polydimethylsiloxane (PDMS) channels are then made via replica molding by
heating a PDMS:curing agent (9:1) mixture on top of the SU-8 master mold. The
channels are then irreversibly bonded to a microscope slide by using a plasma
cleaner (Harrick Plasma, UK) at 18W power for 1 minute. All devices
have a nominal depth of 50 $\upmu$m and a total channel width of \textit{w} =
400$\,\upmu$m. 
For the single junction chip,
the width of the main channel and the inner inlet channel are $w=400\,\upmu$m and $w_{i}=200\,\upmu$m, respectively, whereas the 
width of the outer inlet channels are $(w-w_i)/2=100\,\upmu$m.
For the double-junction device, the width of the main channel is $w=400\,\upmu$m.
For the downstream junction, the width of the inner channel and
outer inlet channels are $w_{m} = 250\,\upmu$m and $(w-w_m)/2=75\,\upmu$m, respectively.
For the upstream junction, the width of the inner and outer inlet channels
are $w_{i} = 100\,\upmu$m and $(w_m-w_i)/2=75\,\upmu$m, respectively.
Aqueous solutions are injected into the device inlets at a constant flow rate
of 3.65 $\upmu$L/min by means of syringe pumps (Harvard, USA). 
To generate a steady state salt gradient, a flow switch valve 
is used to swap the inner stream from a low ($\cL$) to high ($\cH$) salt
concentration solution.

\subsection{Nanoparticle sample preparation~~\label{subsec:sample prep}} 
Carboxylate polystyrene particles ($2\%$ solids) 
were diluted in a 0.1 mM LiCl solution at a particle
concentration of 0.002\% v/v for the $1\,\upmu$m particles and $0.02\%$ v/v for
particles of other diameters.  Liposomes were prepared using standard thin film
hydration~\cite{bangham1967} and subsequent extrusion techniques
\cite{hope1985ex}. Briefly, lipids and lipophlic dye (DiO$_{c_{18}}$) were
dissolved in chloroform and mixed at appropriate ratios. The chloroform was
evaporated using a constant flow of N$_{2}$. The dried lipid film was desiccated
for a minimum of 3 hours before hydration and dilution using a 0.01 mM LiCl +
5$\upmu$M EDTA + 5$\upmu$M HEPES solution, pH = 8.15. The liposome suspension
was then vortexed at 1200 rpm for 60s and extruded 21 times through a 200 nm
polycarbonate membrane filter (Avanti Polar Lipids, US) to maintain an average
polydispersity index (PDI) of 0.1.

\subsection{Image acquisition and processing~~\label{subsec:processing}}

An inverted optical microscope (TE300, Nikon) equipped with a x10 objective
lens (0.25 NA) is used to image particle dynamics at different distances from the
junction.  A fluorescent lamp (CoolLED pE300) is used to excite the sample
which allows the collection of fluorescent intensity data via a CCD camera
(Ximea MQ013MG-ON). Collected data is in the form of 1264 x 1016 px, 16 bit
TIFF images. The depth of field  of the microscope -- i.e. the thickness of the slice region that is in acceptably
sharp focus in the micrographs -- can be calculated as~\cite{inoue1995microscopes} $n_{ind}
\, \lambda_{em} / NA^2 + n_{ind}\,e/(M\cdot NA)\simeq 10~\upmu$m, where $n_{ind}=1$ is the
refractive index of the objective immersion medium (air), $\lambda_{em}=510$~nm
is the nanoparticle emission wavelength, $e=4.8\,\upmu$m is the pixel pitch of
the CMOS camera, $NA=0.25$ and $M=10$ are the objective numerical aperture and
magnification, respectively. The epi-fluorescence micrographs were acquired by
positioning the focal plane nearby either the bottom (glass) wall, the
mid-depth point or the top (PDMS) wall of the device. The location of the
focal plane was established by focusing on colloidal particles permanently
stuck on the bottom and top walls of the microchannel.

A PicoQuant MicroTime 200 time-resolved confocal microscopy platform,
built around an Olympus IX 73 microscope, was used to acquire confocal images in
the x-y plane, as shown in~\figurename~\ref{fig:Fig1}. A plan N 40x water immersion 
objective lens (0.65 NA) was used to take 1024 x 1024 px,
32 bit TIFF images. A monodirectional scanning pattern was used with a learning time of 5 seconds and a dwell time of 2 seconds.
Fluorescent intensity profiles were normalized by first subtracting the
background noise and then dividing by the average intensity of the
bulk obtained from a no salinity gradient configuration. All fluorescent image shown in figures were processed using ImageJ (contrast enhancement, LUT color change).

\subsection{Particle characterization~~\label{subsec:characterisation}}

All size and zeta potential measurements were performed using a ZetaSizer nano
ZS (Malvern Panalytical) at 25\textdegree C. All samples were analyzed a
minimum of 3 times using the monodomal measurement mode for a monodisperse
single population of particles. 
The instrument provided the zeta potential values calculated according to the Smoluchowski’s theory for which
the Debye length is much smaller than the particle size. The zeta potential
values were corrected to account for the finite size of the Debye length
according to Henry’s model\cite{hunter2001foundations}.

\subsection{Numerical methods}
Numerical simulations were performed in Comsol Multiphysics according to the
procedures detailed in our previous work~\cite{singh2020reversible}.  Briefly,
the 3D computational domain consisted in a rectangular channel of width
$2w_i=400\:\upmu$m, depth $h=52\:\upmu$m and length $25 w_i + 5h$. The
hydrodynamic velocity, $\vecu$, pressure $p$, salt concentration $c$ and
particle concentration $n$ were calculated by solving the steady-state
Navier-Stokes equation and the advection-diffusion equations for $c$ and $n$.
At the channel inlet, the boundary condition for the velocity field was
$\vecu=\vecin$, with $\vecin$ the fully developed velocity field at a cross
section of the rectangular channel perpendicular to the flow direction and with
average velocity $U_0$.  The boundary conditions at the channel inlet for the
salt and particle concentration fields are $c=\cH$ and $n=0$ for the outer flow
region (i.e., $|x|>w_i/2$) and $c=\cL$ and $n=n_0$ for the inner flow region (i.e., $|x|\le w_i/2$). At the channel outlet,
the zero normal gradient boundary condition for the pressure, salt and particle
concentrations were imposed. At the remaining walls, the slip boundary
condition $\vecu=-\Gammado \Nab (\ln c) $ was applied together with the zero
flux condition for the salt and particle concentration fields.  The channel
outlet was located at 5 times the channel depth $h$ from the cross section
$z/w_i=25$ to ensure that the boundary conditions at the channel outlet do not
affect the fields near that section.
The particle diffusiophoresis coefficient $\Gammadp=291\,\upmu$m$^2$/s was calculated according
to the procedure detailed in our previous work~\cite{singh2020reversible},
 where the formula provided by Prieve and co-workers\cite{prieve1984motion} was used to
account for the particle size effect on $\Gammadp$. 
A diffusioosmosis coefficient of $\Gammado=1165\,\upmu$m$^2$/s was chosen for the channel walls to
obtain a good quantitative match between experimental results and numerical predictions.

\begin{acknowledgement}
This research was supported by the EPSRC (EP/S013865/1).

\end{acknowledgement}

\begin{suppinfo}

The supplemental file details the results of (i) the control experiments in a single
$\Psi$-junction device without salt concentration gradient, (ii) the experimental test
in a single $\Psi$-junction device with colloidal particles dispersed in the
higher salt concentration stream.

\end{suppinfo}

\bibliography{main_article}

\end{document}


\clearpage
\newpage

\section{Supplementary Information}

\subsection{Control experiment in a single $\Psi$-junction chip}

\begin{figure}[b!]
\begin{center}
\resizebox{6.5in}{!}{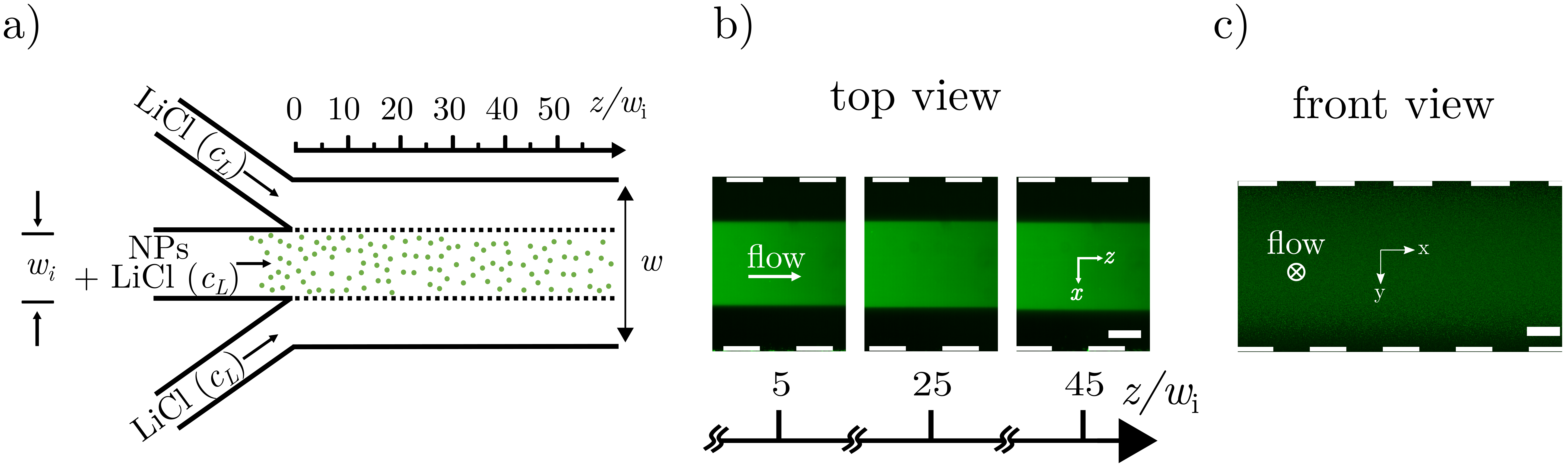}
\end{center}
\caption{Particle dynamics under no salinity gradient conditions $\nabla$c =
0. (a) Schematic diagram of a single $\Psi$-junction microchip, where low salt
concentration is injected in all channels with carboxylate polystyrene nanoparticles (NPs) present in the
central channel only.
(b) Top view epi-fluorescence images taken at different
distances downstream the junction for the flow configuration shown in (a), scale bar =
75 $\mu$m. (c) Confocal image taken at $z/w_{i}$ = 25 for the flow configuration shown
in (a), scale bar = 10 $\mu$m. 
\label{fig:FigS1}
} \end{figure}

A control experiment with no salt concentration gradient 
was conducted by using the same $\Psi$-junction microchip in the main text.
A colloidal suspension at low salt concentration ($c_{\sst L}$) was injected in the inner channel, whereas
an aqueous solution with identical level of salinity ($c_{\sst L}$) was pumped 
in the outer channels (Fig.\,\ref{fig:FigS1}a). 
In the absence of a chemical gradient, the colloids remained homogeneously distributed
within the inner region of the channel and the colloidal band maintained a constant width downstream
the junction (Fig.\,\ref{fig:FigS1}b). 
No peak formation and subsequent convergence were
visible and the particle distribution profile along the channel depth
remained uniform (Fig.\,\ref{fig:FigS1}c). 
This is because no salinity gradient exists in
any direction. Consequently,  electrokinetic phenomena such as
diffusiophoresis and diffusiosmosis do not occur in the system. 
Furthermore, the flow is characterized by
a high Peclet number, $Pe=U_0\,w /D_c\simeq10^6$, where $D_c$ is the particle diffusivity given by
$D_c = k_b T / 6\pi\mu a=2.4\times10^{-12}\,$m$^2$/s with $\mu=0.9\times10^{-3}\,$Pa~s the solution viscosity
and $a$nm the particle radius. Therefore
particle diffusion is negligible compared to convective transport,
and since both inertia and gravity effects are also negligible, the colloids behave as passive tracers.

\subsection{Salinity conditions of NP solution for peak formation}
\begin{figure}[b!]
\begin{center}
\resizebox{6.5in}{!}{\input{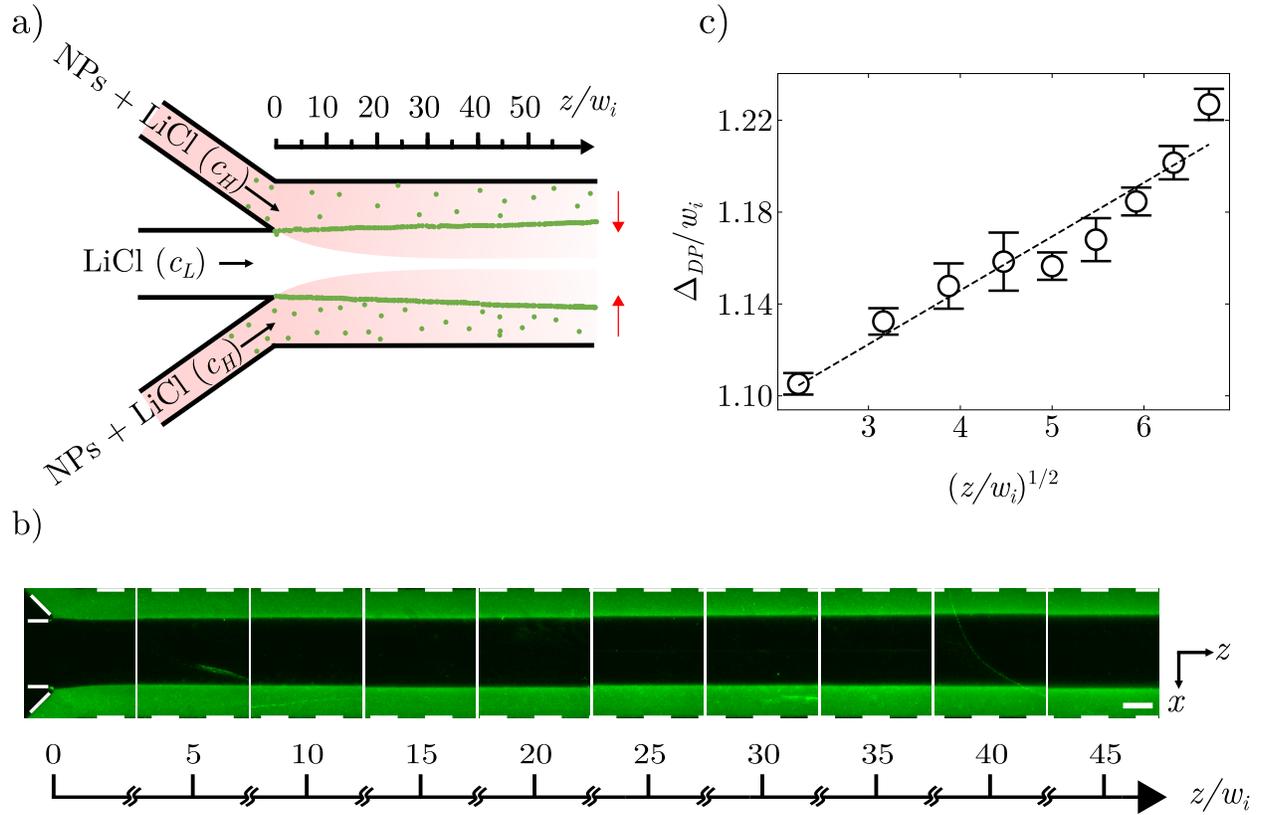}}
\end{center}
\caption{Absence of the hereby reported focusing mechanisms when colloids are in high
salinity. (a) Schematic diagram of a single $\Psi$-junction microchip, where
low salt is injected in the inner channel and high salt with colloids being
injected in the outer channels. (b) Epi-fluorescence images taken at different
distances downstream the junction  for the flow configuration in (a), scale bar
= 75 $\mu$m. (c) The dynamics of particle migration against
the square root of the longitudinal distance downstream the junction.
\label{fig:FigS2}} \end{figure}

We find that peak formation due to the reported focusing effect does not occur
when the carboxylate polystyrene nanoparticles are dispersed in the high salt ($c_{\sst H}=10~$mM)
solution in the outer channels and not in the low salt ($c_{\sst L}=0.1~$mM) solution in
the inner channel (Fig.\,\ref{fig:FigS2}a). Consequently,
this led to the development of the two $\Psi$-junction microchip for
nanoparticle fractioning to target advanced applications. Moreover, we observe
a previously reported particle accumulation effect that has been outlined in
the main text. Fig.\,\ref{fig:FigS2}c substantiates this, as accumulated particles diverge
towards higher salinity regions at a $\sqrt{z}$ trend downstream the junction.

\clearpage
\newpage